 \documentclass[aip,jcp,reprint,twocolumn,superscriptaddress,floatfix,nofootinbib]{revtex4}

\usepackage{graphicx}
\usepackage{color}
\usepackage{amsmath}
\usepackage{amssymb}
\usepackage[english]{babel}
\usepackage[utf8]{inputenc}
\usepackage{bm}
\usepackage{url}
\usepackage{ulem}
\usepackage[colorlinks=true,linkcolor=blue,citecolor=blue,urlcolor=blue]{hyperref}

\definecolor{dgreen}{rgb}{0.0,0.7,0.0}
\definecolor{purple}{rgb}{0.6,0.0,0.6}

\begin{document}
%
% TITLE
%
\title{Unexpected dipole instabilities in small molecules after ultrafast XUV irradiation}
\author{P.-G.~Reinhard}
\affiliation{Institut f\"ur Theoretische Physik, Universit\"at Erlangen,
             Erlangen, Germany}
\author{D. Dundas}
\affiliation{School of Mathematics and Physics, Queen's University Belfast}
\author{P. M. Dinh}
\affiliation{Laboratoire de Physique Théorique, Université de Toulouse, CNRS, UPS, France}
\author{M.~Vincendon}
\affiliation{Laboratoire de Physique Théorique, Université de Toulouse, CNRS, UPS, France}
\author{E.~Suraud}
\affiliation{Laboratoire de Physique Théorique, Université de Toulouse, CNRS, UPS, France}
\date{\today}
\pacs{                                      }
\begin{abstract}
We investigate the depletion of single-electron states in small molecules under the
influence of very short XUV pulses. 
In N$_2$, for a
certain window of XUV energies around 50 eV, we observe a marked occupation
inversion, i.e. a situation where depletion of the deepest bound valence
electron state is much larger than for any other state. This represents a realistic
mechanism which is able to cut, almost instantaneously,
a hole into a deep lying state, 
a situation which is often assumed ad
hoc in numerous theoretical studies of energetic ultrafast  processes.
This occupation inversion furthermore drives a
dipole instability, i.e. a spontaneous reappearance of the dipole signal
long after the laser pulse is over and the dipole signal has died out. The dipole signal  
that emerges from this instability can be identified as a particular low-energy structure in
photo-electron spectra.
\end{abstract}
\maketitle

With the advent of ultrashort laser pulses in the XUV to
X-ray frequency regime, attoscience has become a dynamically growing field over the last
two decades~\cite{Kra09}. The short pulse durations and the great flexibility 
in shaping pulses nourish hope for
time-resolved measurements down to the electronic timescale
\cite{Cal16}. The
typically high photon energies extend electron excitations to deeper lying states
often down to core states.  This
can trigger various relaxation processes as Auger decays~\cite{Ued19}, 
interatomic (or intermolecular) Coulomb
decays~\cite{Jah20}, charge migration in covalent
molecules~\cite{Mar19,Mar20} or giant autoionization resonance in high
harmonic generation~\cite{Wah19,Aus21}.  A fully detailed theoretical
treatment of ultrafast excitation
from the ground state to
high-energy configurations remains  highly demanding 
\cite{Rub19}. 
One thus often
ignores the detailed excitation process by assuming that an attosecond
XUV pulse instantaneously removes a bound
electron~\cite{Ced86,Ced99}. This considerably simplifies the picture
since the excitation process is not considered explicitly and one can choose
at will where the hole is created.  Such an excitation model is thus routinely 
used to explore the dynamics of charge migration and charge
transfer in molecular systems
\cite{Wei96,Rem99,Rem06,Kul11,Kul16,Gol21,Gon21,Kha21}.

We take here a closer look at the detailed excitation
process induced by such a short XUV pulse. We simulate the
full dynamical interaction of an ultrashort XUV pulse with a
multi-electronic system, and the electronic emission thereof, 
using Time-Dependent Density Functional Theory (TDDFT) \cite{Run84}. 
Thanks to a pseudopotential with frozen cores
\cite{Goe96}, we only consider valence electrons and focus
on the depletion of the lowest occupied valence
state, easily attainable with an XUV laser. 
Indeed low-frequency pulses skim electrons from the Fermi
surface while deeper lying electrons come increasingly into play with
increasing pulse frequency \cite{Vid10a,Din12b,Wop15b}. 
One might expect an equi-distribution of depletion over all valence states, as observed in 
Na clusters~\cite{Vid10a,Din12b}
and C$_{60}~$\cite{Wop15b}. 
However,  when  considering  ultrafast 
pulses in the few femtosecond
regime,  we  observe  in  covalent   molecules
an ionization mechanism very close to the creation of an instantaneous deep-hole.
A remarkable consequence is that
this excitation leads to a dipole instability, with a delayed reappearance of the dipole signal well after
the pulse  is over and the dipole signal has died out.
This instability leaves as an experimentally accessible footprint
a low-energy peak in photoelectron spectra.  We only consider here the simple 
N$_2$ molecule, but  we found similar behavior in other small molecules such as 
acetylene. Additionally, our  results are robust when 
including ionic motion and incoherent electronic
dynamical correlations within a Relaxation Time
Approximation (RTA)~\cite{Rei15d}. This partly answers the question of the validity of TDLDA 
for such ultrafast setups.

We describe the electrons 
with the widely used  TDDFT \cite{Run84}, at the 
level of the Local Density Approximation (LDA) \cite{Dre90}
(Time-Dependent LDA (TDLDA)), using the
 functional of  \cite{Per92}.
 % We refer to this approach as the 
%Time-Dependent LDA (TDLDA). 
The single particle (s.p.) wave functions
are discretized on a real-space grid.  The dynamical treatment of
ionization requires realistic s.p. energies, particularly for 
the HOMO. We thus add  a 
Self-Interaction Correction (SIC) \cite{Per81}, actually the
simple and efficient Average-Density SIC (ADSIC)
\cite{Fer34,Leg02a,Klu13a,Rei21}. 
Electron emission is evaluated through absorbing 
boundary conditions via  a mask function \cite{Ull00b,Rei06c}. The coupling
between electrons and ions is described using
Goedecker type pseudopotentials
\cite{Goe96}.
The ionic positions are kept frozen
since the time scale of the electronic processes considered is very
short. 
We have also 
performed calculations including electronic dynamical 
correlations at the level of the RTA
\cite{Rei15d,Din18,Din22aR}. 
%A detailed description of the theory,
%the numerical realization in coordinate-space representation, and the
%actual code QDD (Quantum Dissipative Dynamics) is found in
%\cite{Din22aR}. 
As our test case will be  N$_2$ which is an axially symmetric system, %($z$ axis), 
we employ here the
2D axial version of the general code QDD (Quantum Dissipative Dynamics) \cite{Din22aR}. It
allows us to scan a larger Hilbert space for the  (initially unoccupied) states used in 
RTA dynamics \cite{Rei15d,Din22aR}.
Wave functions
and fields are represented on a 2D axial grid with a grid spacing of
0.25 a$_0$: 301 grid points in the axial ($z$) direction, and 151 
in the radial direction.  The distance between the ions is taken as the
minimum of the Born-Oppenheimer surface which is 2.02 a$_0$ for our setup
(pseudopotential and ADSIC).
The stationary electronic state is
computed using an accelerated gradient method and the time-dependent
Kohn-Sham equations are propagated using a time-splitting technique
\cite{Cal00,Din22aR} with a time step of  0.0048 fs.

We will consider three observables.  The first one is the time
evolution of the dipole moment along laser polarization/molecular axis  $z$: 
%\begin{equation}
%  \mathbf{d}(t)
$D(t)  =
%  \int \textrm d^3r\,e\mathbf{r}\,\sum_{\alpha=1}^N|\varphi_\alpha(\mathbf{r},t)|^2
  \int \textrm d^3 \mathbf r\, e z \, \varrho(\mathbf{r}).
%  \sum_{\alpha=1}^N|\varphi_\alpha(\mathbf{r},t)|^2
$
%\end{equation}
The single electron density is computed as $\varrho(\mathbf{r}) = \sum_{\alpha} w_\alpha(t)  |\varphi_\alpha(\mathbf{r},t)|^2$ 
where $\varphi_\alpha(\mathbf{r},t)$ is the time-dependent s.p. wave function of state $\alpha$ and $w_\alpha(t)$ its occupation number.
The summation here runs on all computed states. At  LDA  level, 
only occupied states are treated and for these we have
$w_\alpha=1$ and independent of time. 
At RTA level, the $w_\alpha$'s become fractional and time-dependent \cite{Rei15d,Din22aR}.
The dipole signal, recorded from initializing the dynamics
with a small kick, provides  the optical response 
of the system by spectral analysis 
\cite{Cal97b}.

The second observable is the electron content  $n_\alpha$ per s.p. wave function,
and its complement, the depletion $\bar{n}_\alpha$:
%, which
%are computed as:
\begin{equation}
  {n}_\alpha(t)
  =
  w_\alpha\int \textrm d^3 \mathbf r\,|\varphi_\alpha(\mathbf{r},t)|^2
  \quad,\quad
  \overline{n}_\alpha(t)
  =
  1-n_\alpha(t)
  \;.
\label{eq:deplet}
\end{equation}
The  total number of escaped electrons is 
$N_\mathrm{esc}(t)=\sum_\alpha\overline{n}_\alpha(t)=N-\sum_\alpha n_\alpha(t)$, where $N$ is the initial 
 number of computed electrons (10 for N$_2$).

Lastly, we analyze the electronic emission dynamics through
Photo-Electron Spectra (PES). 
At several measuring 
points shortly before the absorbing
boundaries begin, we record the time evolution of each s.p.
wave function
and finally Fourier transform
them to the frequency domain. This  provides the 
spectrum of
kinetic energies of the escaping electrons, that is the PES
\cite{Poh00,Poh01,Wop15aR}.  In  strong fields an
additional phase correction has to be added~\cite{Din13c}.

The system is excited by an XUV pulse modeled as a classical
(coherent) photon field linearly polarized along the symmetry axis
$z$. The corresponding potential reads:
%.  It is described by the
%following time-dependent Coulomb potential:\DDfoot{\protect $\mathbf{z}$ in Eq. (3a), rather than 
%  $\hat{\mathbf{z}}$?}
\begin{subequations}
\begin{eqnarray}
  V_\mathrm{pulse}
  &=&
%  e{\cal\bf E}_0f(t)\!\cdot\!\hat{\mathbf{z}}
  e{E}_0z\,f(t)
  \cos \big[ \omega_\mathrm{XUV}(t-T_\mathrm{pulse}) \big]
\label{eq:laserr}
\\
  f(t)
  &=&
  \left\{\begin{array}{ll}
    \sin^2{\left(\displaystyle \pi\frac{t}{2T_\mathrm{pulse}}\right)}
        &\mbox{$t\in\{0,\, 2T_\mathrm{pulse}\}$}  \\
    0   &\mbox{otherwise}
  \end{array}\right.
\label{eq:sinpulse}
\end{eqnarray}
\end{subequations}
The pulse parameters are  the frequency $\omega_{\rm XUV}$, the duration
$T_\mathrm{pulse}$, and the field strength ${E}_0$ (linked to the
pulse intensity as  $I \propto {E}_0^2$). 
We use
$T_\mathrm{pulse}=1$ fs and, for a given $\omega_{\rm XUV}$, 
adjust $E_0$ so that the total ionization $N_\mathrm{esc}$  levels off 
asymptotically at
about 1.

%\section{Results on the N$_2$ molecule}
%\label{sec:results}

%As a first step, t
The electronic ground state of N$_2$ with its ten valence electrons is prepared by solving
the static Kohn-Sham equations.  The emerging occupied ground state s.p. states, each doubly occupied,
have the following energies: $\varepsilon_1=-32.5$
eV, $\varepsilon_2=-18.6$ eV, $\varepsilon_3=\varepsilon_4=-17.7$ eV,
$\varepsilon_5=\varepsilon_\mathrm{HOMO}=-15.1$ eV ($=-$ Ionization Potential, IP).  They compare well with 
the experimental values for the last three levels,
namely $-18.6$ eV, $-16.6$ eV, and $-15.5$ eV, respectively.

\begin{figure}[htbp!]
\centerline{\includegraphics[width=\linewidth]{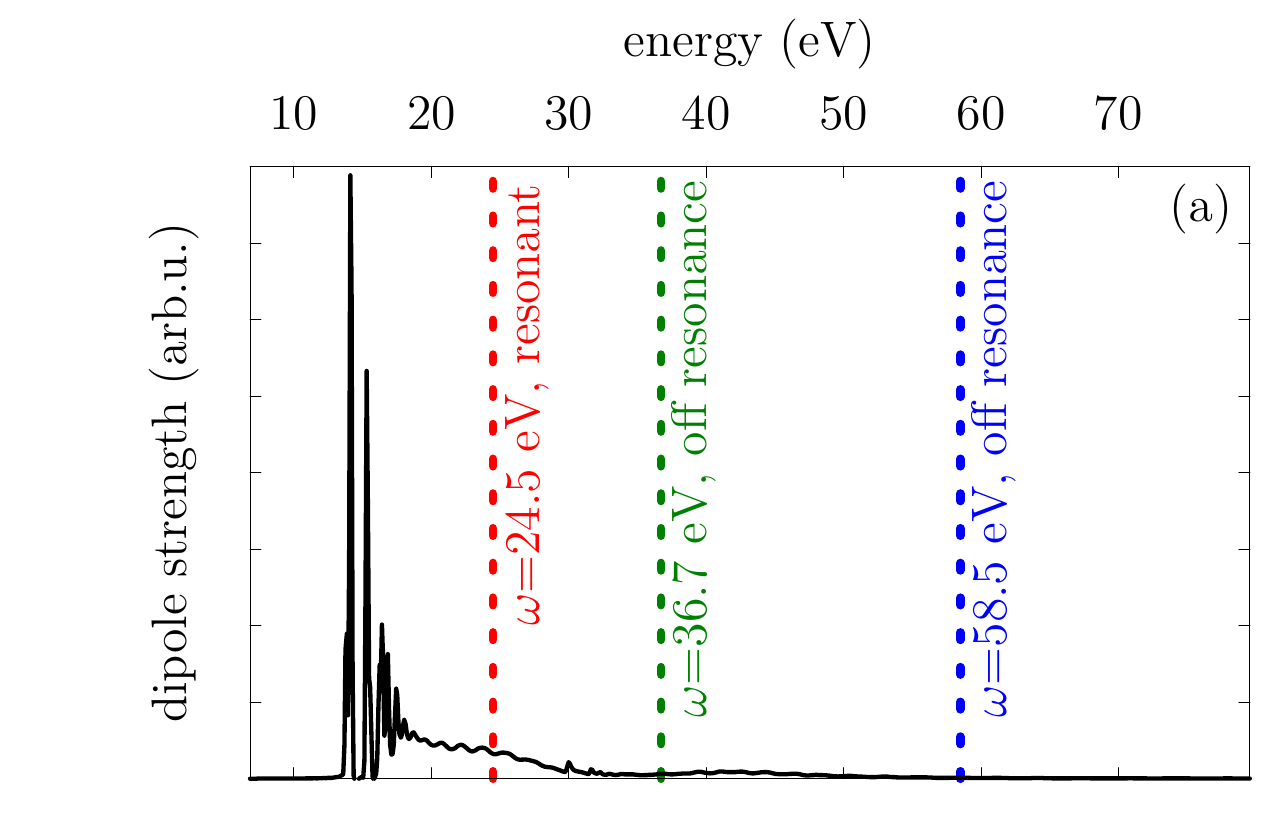}}
\centerline{\includegraphics[width=\linewidth]{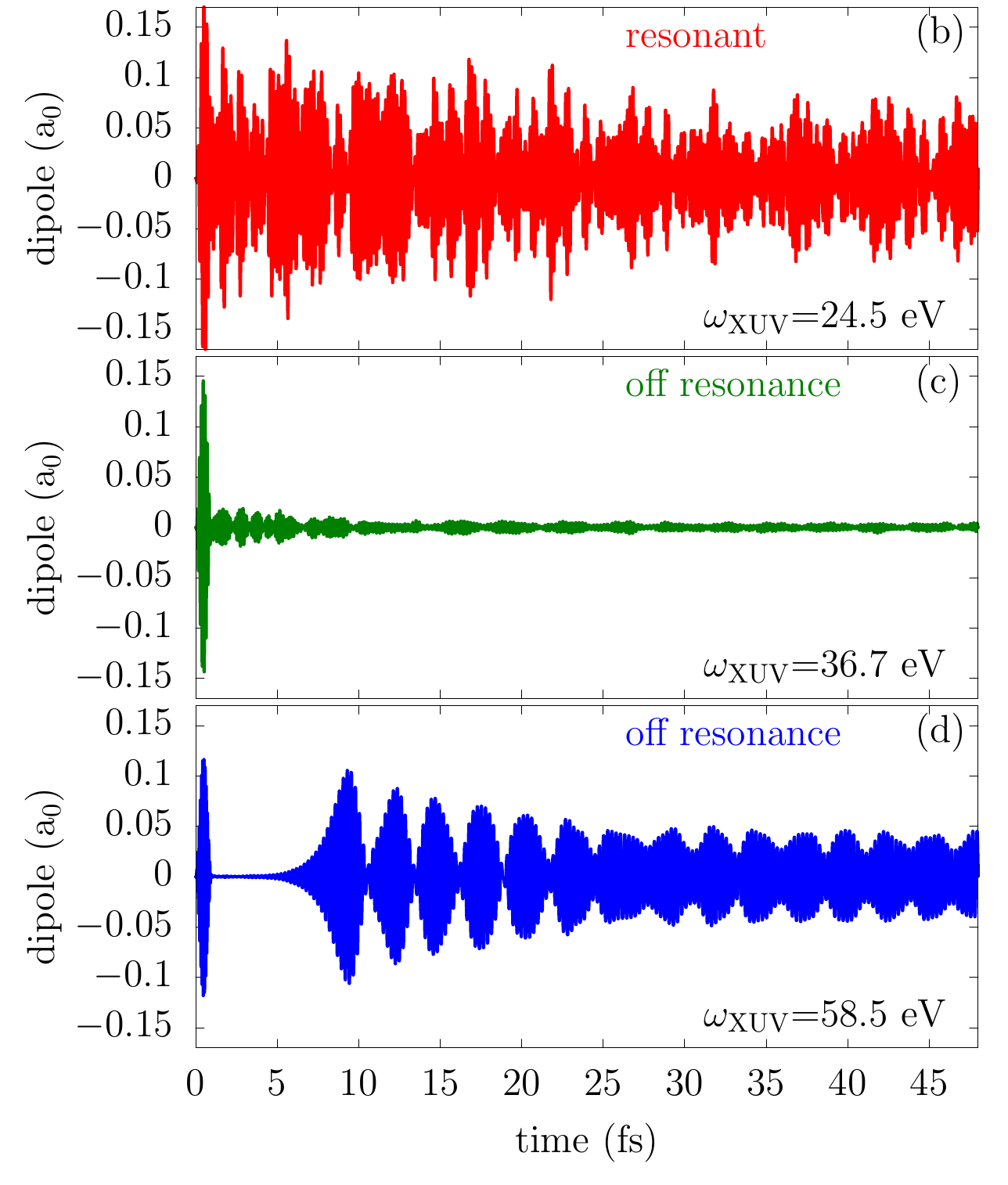}}
\caption{Panel (a): Optical response of N$_2$ (linear  scale). The vertical
  dashes indicate the three  frequencies considered in panels (b), (c) and (d). 
Panels (b) to (d): dipole moment
after   an
XUV excitation of duration of 1 fs, frequency
$\omega_\mathrm{XUV}$ as indicated, and field strength $E_0$ such that
the total ionization is about one charge unit for each case
which is $I=0.56\times 10^{15}$W/cm$^2$ (panel b), $I=1.4\times10^{15}$W/cm$^2$ (panel c), and
$I=7.0\times10^{15}$W/cm$^2$  (panel d).
}
\label{fig:spectr-dipole}
\end{figure}
We now turn to the dynamical response of the molecule.
We first consider the optical response of N$_2$
in panel (a) of Fig.~\ref{fig:spectr-dipole}.  In the lower energy region
below and near emission threshold (IP=15.1 eV), we see a much
fragmented spectrum, typical for covalent molecules.
At higher energies, it
turns into a smooth and considerably spread continuum spectrum. In panels (b)--(d),
we consider the time evolution of the dipole moment for three selected laser frequencies,
that is 24.5, 36.7 and 58.5 eV, indicated in panel (a), which deliver different 
characteristic behaviors.
The first value lies 
in the upper tail of a peak in the optical response. 
We thus expect a resonant response. 
The two other values lie in regions where the optical response vanishes and we
expect non-resonant dynamics.
This is indeed confirmed in panels (b)--(d) of Fig.~\ref{fig:spectr-dipole}.
In panel (b), sizeable
dipole oscillations persist long after the pulse duration, 
typical of a resonant excitation. Panel (c) instead shows a standard
off-resonant response: the dipole signal quickly dies off once
the pulse is over. In the largest frequency case shown in panel (d),
the dipole moment initially behaves as in a non-resonant case, i.e. dying
out as expected. However, surprisingly, a longlasting 
dipole signal reappears after $\simeq 5$ fs. Closer inspection of the evolving dipole on
a logarithmic scale (not shown) reveals that the envelope of the
dipole signal up to $\simeq 8$ fs increases exponentially, as is typical for
an instability.  In order not to be fooled by artifacts, we have
scrutinized our numerical treatment by varying numerical parameters and by
performing control calculations using two other full 3D 
computer packages (EDAMAME~\cite{dundas:2012} and QDD~\cite{Din22aR}). The effect persists.  
It is thus likely to be genuine, at least at TDLDA level, and deserves a further analysis.

\begin{figure}[htbp!]
\centerline{\includegraphics[width=\linewidth]{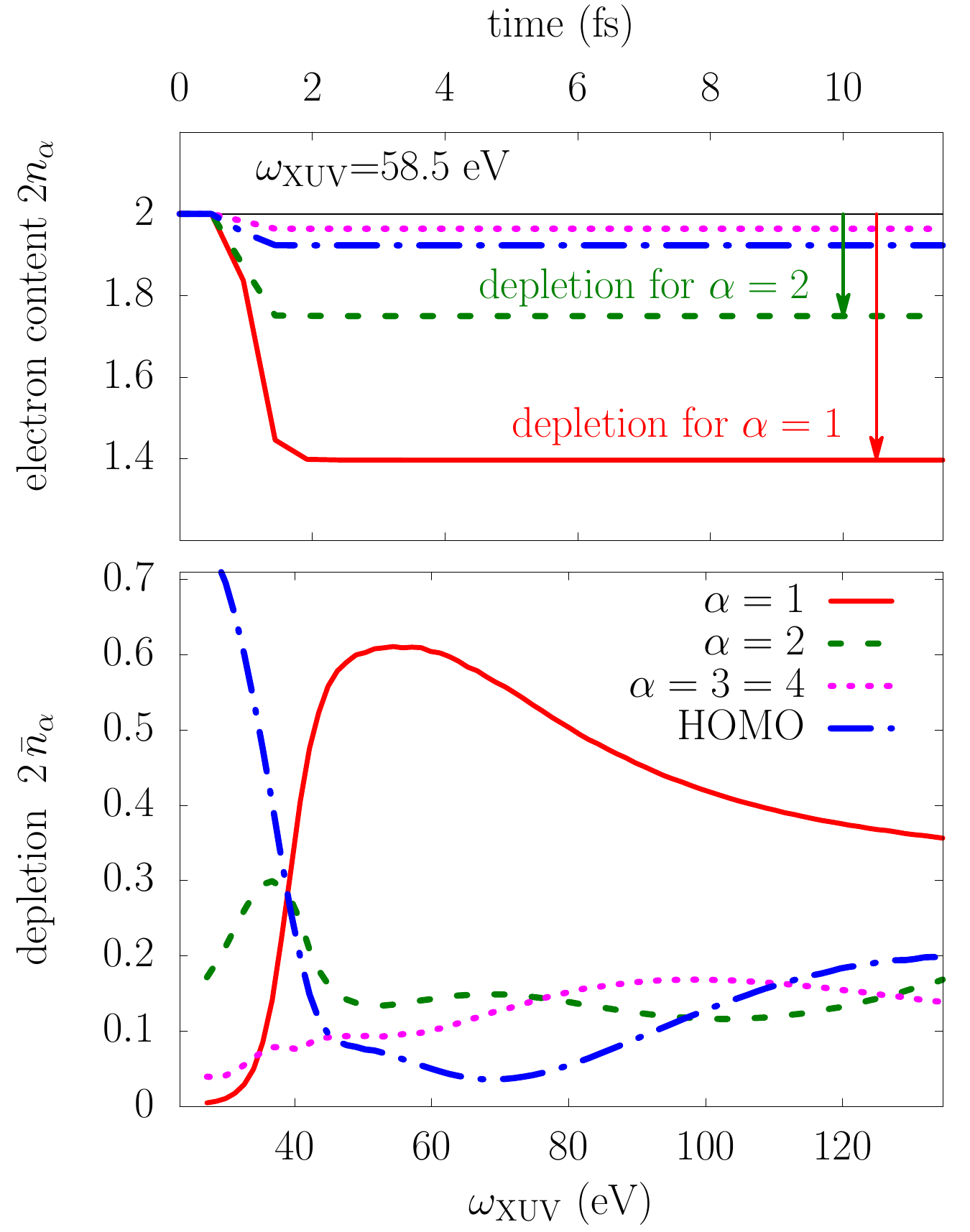}}
\caption{Top: Time evolution of the electron content 
  $2n_\alpha(t)$ (factor 2  for spin degeneracy) of each
  s.p. state $\alpha$ (see legend in lower panel), after a 1~fs XUV pulse of frequency
  $\omega_\mathrm{XUV}=58.5$ eV. 
  Level depletions $\overline{n}_1$ and $\overline{n}_2$ are indicated by  vertical arrows.
Bottom: Level depletion $2\overline{n}_\alpha$ of the five initially
  occupied states as functions of frequency $\omega_{\rm XUV}$. Field strengths $E_0$
  are adjusted to  total ionizations of about 1  for each case.}
\label{fig:depletion}
\end{figure}
In Fig.~\ref{fig:depletion}, we first analyze level depletions, see
Eq.~(\ref{eq:deplet}). 
The upper panel
illustrates the time evolution of electron content for one chosen
pulse frequency. 
Electron emission is very fast and the electron
content $n_\alpha$ and the depletion $\overline{n}_\alpha$ accordingly level off very quickly.
The exciting aspect is that the lowest level has the largest depletion, 
much larger than the depletion for all the other (higher
lying) states. This comes very close to the instantaneous-hole
scenario used in many theoretical investigations.

The lower panel of Fig.~\ref{fig:depletion} displays the final
depletions as functions of pulse frequency $\omega_{\rm XUV}$. At the lower frequencies,
electron depletion occurs predominantly from the HOMO orbital, as expected. 
At the upper end, depletion
approaches an equi-distrbiution similar to what we observed  for
Na clusters and C$_{60}$ \cite{Vid10a,Din12b,Wop15b}. But in between,
we find an occupation inversion  over a wide range of frequencies
with, by far, the dominant depletion coming from the deepest level $\alpha = 1$.
We have observed the same effect in other covalent 
molecules such as acetylene.
Such an occupation inversion in
a certain frequency range might be  due to a continuum
resonance enhancing  the transition matrix elements
selectively, but this requires  a closer inspection.

%It is also obvious that there must be a relation 
The link between occupation
inversion and the observed dipole instability looks striking. In a simple model
with a couple of fixed electron levels coupled to the photon
field, one can show that the mechanism looks  the same as in a
laser \cite{Hak83aB,Hak91aB}. The energy reservoir contained in the
occupation inversion feeds the dipole oscillations coherently. This leads 
to an initial exponential increase of the dipole amplitude which
turns into a steady oscillation once the reservoir is used up.  For reason
of space, we will present the details in a forthcoming publication. 

%\ES{The occurence of the long time dipole instability might thus be understood that way. We have 
%already eliminated spurious computational problems. 
For the moment, there remain two questions yet to be addressed:
firstly, to propose a potential experimental observation of such an
effect, and secondly, to exclude any artifact from the mean-field
approach in TDLDA.  Let us successively explore both aspects.

\begin{figure}[htbp!]
\centerline{\includegraphics[width=\linewidth]{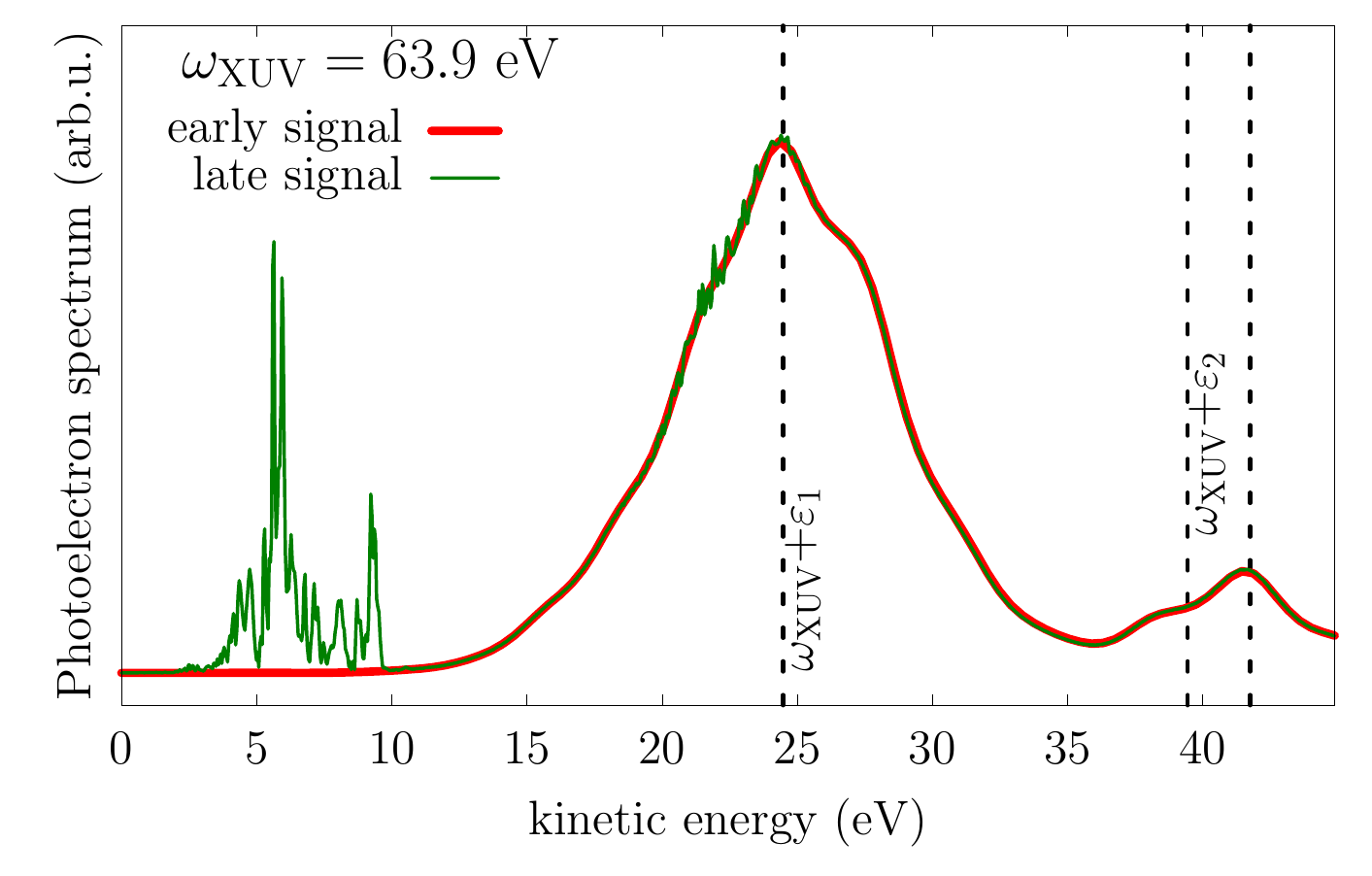}}
\caption{PES {(linear scale)}, after a 1 fs XUV excitation with $\omega_\mathrm{XUV}=63.9$ eV,
  and field strength 
  $E_0=15.9$ eV/a$_0$, computed at 10 fs  (thick green line) or at 100 fs (thin black line).
  The faint dashed vertical lines indicate the 
  s.p. energies shifted by the pulse frequency, see text for detail.}
\label{fig:PES}
\end{figure}
Depletion and dipole moment are not directly
observable by experiment. Still, a  possible experimental identification of
the dipole instability can be found in the PES.  
Figure~\ref{fig:PES} shows the PES  after irradiation by a 1fs pulse with
$\omega_\mathrm{XUV}=63.9$ eV
and $E_0=15.9$ eV/a$_0$.  
Two different PES are plotted: one
recorded in an early time window up to 10 fs (basically before the onset of the instability) and  one computed
at the end of the simulation time (100 fs). The early PES
exhibits the standard peaks for a one-photon process with 
energies at $\varepsilon_\alpha +  \omega_{\rm XUV}$ 
(vertical dashes in Fig. \ref{fig:PES}).
These peaks are very broad because the pulse is very
short and because the s.p. energies move down 
by about 8~eV with respect 
to the ground state s.p. energies, due to emission and subsequent
Coulomb charging \cite{Wop15aR}. 
The late PES consists of the early PES plus an additional low energy structure, which
does not fit to any combination
$\varepsilon_\alpha + n \omega_{\rm XUV}$ (with integer $n$). 
The fact that this structure shows 
up at later times suggests that it is  a signature of the dipole
instability.
%~\DDfoot{I guess these spectra are obtained by applying some sort
%of window to the collected signal. Would it be possible to scan the data with a moving window to produce a
%short-time Fourier transform spectrogram, similar to what people do for HHG to see clearly when the low energy
%structure first appears?}. 
Indeed, the Fourier transform of the dipole moment signal at late times exhibits an oscillation frequency
of $\omega_{\rm instab}=16.4$ eV. And the peaks at low energies might
correspond to the actual s.p. energies (IP$\approx 28.6$ eV) plus
$2\omega_{\rm instab}$.

Time-resolved PES measurements could thus produce
unambiguous signals of the dipole instability. But 
achieving this remains demanding.
Firstly, one needs cases with  occupation inversion. 
Secondly, 
the PES from one-photon processes alone has to hit  a well-separated, 
energy gap for  the structure associated with the instability to inhabit.
Further theoretical investigations are needed to find more examples of 
promising laser setups.

Thus far, we have considered frozen ions and a pure
mean-field (TDLDA) description of electrons.  
We checked the effect of ionic motion and it makes no difference. This 
is plausible in view of the extremely short time scales and   ion masses.
The impact of dynamical electron-electron
correlations requires a closer look. 
To that end, we include electronic dissipation within RTA \cite{Rei15d,Din22aR}
(incoherent dynamical correlations).
\begin{figure}[htbp!]
\includegraphics[width=\linewidth]{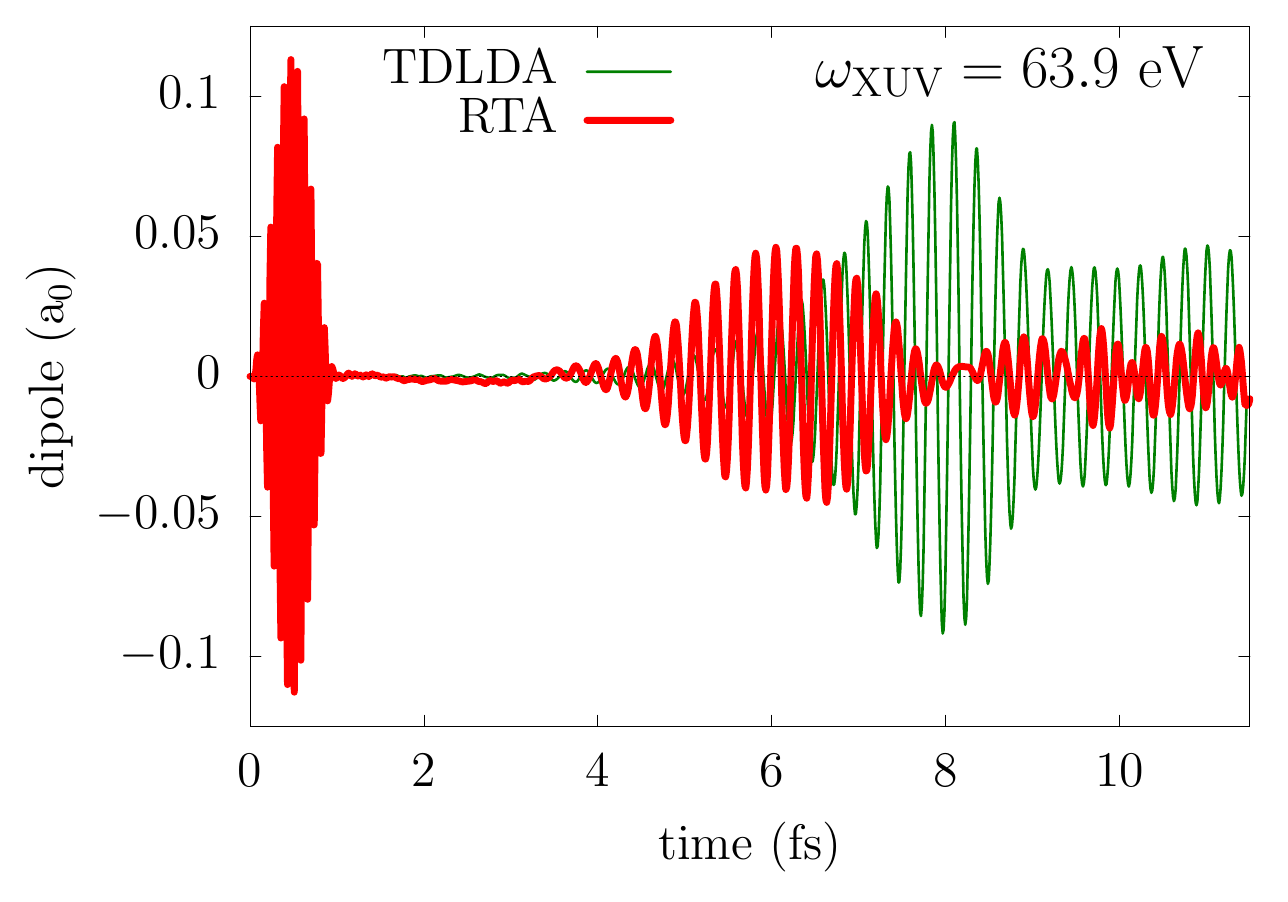}
\caption{Time evolution of the dipole moment 
computed within TDLDA or RTA. Same pulse parameters as in Fig.~\ref{fig:PES}. }
\label{fig:LDARTA}
\end{figure}
Figure~\ref{fig:LDARTA} compares the time evolution of the TDLDA dipole moment to the RTA result. 
The pulse parameters are the same as in Fig.~\ref{fig:PES}. 
Dissipation reduces the dipole instability but a sizeable dipole oscillation remains. 
The result is corroborated by a simple comparison of time scales. The dipole
instability increases one order of magnitude in 2 fs while the typical
relaxation time for N$_2$ at the given excitation energy is about 5 fs.
This estimate of time scales also tells us that the XUV pulses have to be
shorter than the relaxation time, to let the instability
live, and shorter than the instability time to disentangle dipole
oscillations created by the laser pulse and those from the instability.
For the ultrafast scenarios investigated here time scales are thus  forgiving and tend to
validate TDLDA.

To conclude,
we have investigated the excitation dynamics of electrons in N$_2$ irradiated
by an ultrashort XUV pulse. It turns  out
that a proper choice of laser parameters allows to produce an 
almost perfect population inversion. Such a mechanism
represents a physical realization of the instantaneous hole excitation used in 
numerous modelings of ultrafast processes. 
As a very interesting side effect, we find that the
population inversion generates by an unstable dynamical regime of the electron cloud, which 
develops spontaneously sizeable dipole oscillations. 
We checked that the instability is not caused by numerical artifacts
and that it persists even when dynamical electron correlations are included.
A possible experimental signal can be found in photoeletron spectra (PES)
and even better in time-resolved PES.

These results call for  further theoretical investigations which are underway.
Preliminary results indicate that the observed instability is not specific to 
N$_2$; we found similar behaviors in acetylene and  in 1D model systems.
We also observed dipole instabilities with
%much different laser pulses, 
more common longer laser pulses at lower intensities for which 
ionic and RTA time scales are shorter than the onset of instability, 
which then heavily questions the validity of TDLDA in such cases.
Moreover, we can reproduce the dipole instability with instantaneous hole
excitations which simplifies modeling and thus allows us to explore
trends as, e.g., with excitation energy or hole states.
These complementary results will be presented in forthcoming publications.
For subsequent studies remains the question of what a more elaborate theory 
with coherent correlations would produce in such a case.  And, of
  course, a final clarification has to come from experiment.

\vskip 0.5cm
Acknowledgments:
This work was granted
access to the HPC resources of CalMiP (Calcul en Midi-Pyr\'en\'ees)
under the allocation P1238, and of RRZE (Regionales Rechenzentrum
Erlangen).

\newpage
\bibliographystyle{apsrev}
\bibliography{XUV} 

\begin{thebibliography}{45}
\expandafter\ifx\csname natexlab\endcsname\relax\def\natexlab#1{#1}\fi
\expandafter\ifx\csname bibnamefont\endcsname\relax
  \def\bibnamefont#1{#1}\fi
\expandafter\ifx\csname bibfnamefont\endcsname\relax
  \def\bibfnamefont#1{#1}\fi
\expandafter\ifx\csname citenamefont\endcsname\relax
  \def\citenamefont#1{#1}\fi
\expandafter\ifx\csname url\endcsname\relax
  \def\url#1{\texttt{#1}}\fi
\expandafter\ifx\csname urlprefix\endcsname\relax\def\urlprefix{URL }\fi
\providecommand{\bibinfo}[2]{#2}
\providecommand{\eprint}[2][]{\url{#2}}

\bibitem[{\citenamefont{Krausz and Ivanov}(2009)}]{Kra09}
\bibinfo{author}{\bibfnamefont{F.}~\bibnamefont{Krausz}} \bibnamefont{and}
  \bibinfo{author}{\bibfnamefont{M.}~\bibnamefont{Ivanov}},
  \bibinfo{journal}{Rev. Mod. Phys.} \textbf{\bibinfo{volume}{81}},
  \bibinfo{pages}{163} (\bibinfo{year}{2009}).

\bibitem[{\citenamefont{Calegari et~al.}(2016)\citenamefont{Calegari, Sansone,
  Stagira, Vozzi, and Nisoli}}]{Cal16}
\bibinfo{author}{\bibfnamefont{F.}~\bibnamefont{Calegari}},
  \bibinfo{author}{\bibfnamefont{G.}~\bibnamefont{Sansone}},
  \bibinfo{author}{\bibfnamefont{S.}~\bibnamefont{Stagira}},
  \bibinfo{author}{\bibfnamefont{C.}~\bibnamefont{Vozzi}}, \bibnamefont{and}
  \bibinfo{author}{\bibfnamefont{M.}~\bibnamefont{Nisoli}},
  \bibinfo{journal}{J. Phys. B} \textbf{\bibinfo{volume}{49}}
  (\bibinfo{year}{2016}).

\bibitem[{\citenamefont{Ueda et~al.}(2019)\citenamefont{Ueda, Sokell,
  Schippers, Aumayr, Sadeghpour, Burgdörfer, Lemell, Tong, Pfeifer, Calegari
  et~al.}}]{Ued19}
\bibinfo{author}{\bibfnamefont{K.}~\bibnamefont{Ueda}},
  \bibinfo{author}{\bibfnamefont{E.}~\bibnamefont{Sokell}},
  \bibinfo{author}{\bibfnamefont{S.}~\bibnamefont{Schippers}},
  \bibinfo{author}{\bibfnamefont{F.}~\bibnamefont{Aumayr}},
  \bibinfo{author}{\bibfnamefont{H.}~\bibnamefont{Sadeghpour}},
  \bibinfo{author}{\bibfnamefont{J.}~\bibnamefont{Burgdörfer}},
  \bibinfo{author}{\bibfnamefont{C.}~\bibnamefont{Lemell}},
  \bibinfo{author}{\bibfnamefont{X.-M.} \bibnamefont{Tong}},
  \bibinfo{author}{\bibfnamefont{T.}~\bibnamefont{Pfeifer}},
  \bibinfo{author}{\bibfnamefont{F.}~\bibnamefont{Calegari}},
  \bibnamefont{et~al.}, \bibinfo{journal}{J. Phys. B}
  \textbf{\bibinfo{volume}{52}}, \bibinfo{pages}{171001}
  (\bibinfo{year}{2019}).

\bibitem[{\citenamefont{Jahnke et~al.}(2020)\citenamefont{Jahnke, Hergenhahn,
  Winter, D\"orner, Fr\"uhling, Demekhin, Gokhberg, Cederbaum, Ehresmann, Knie
  et~al.}}]{Jah20}
\bibinfo{author}{\bibfnamefont{T.}~\bibnamefont{Jahnke}},
  \bibinfo{author}{\bibfnamefont{U.}~\bibnamefont{Hergenhahn}},
  \bibinfo{author}{\bibfnamefont{B.}~\bibnamefont{Winter}},
  \bibinfo{author}{\bibfnamefont{R.}~\bibnamefont{D\"orner}},
  \bibinfo{author}{\bibfnamefont{U.}~\bibnamefont{Fr\"uhling}},
  \bibinfo{author}{\bibfnamefont{P.~V.} \bibnamefont{Demekhin}},
  \bibinfo{author}{\bibfnamefont{K.}~\bibnamefont{Gokhberg}},
  \bibinfo{author}{\bibfnamefont{L.~S.} \bibnamefont{Cederbaum}},
  \bibinfo{author}{\bibfnamefont{A.}~\bibnamefont{Ehresmann}},
  \bibinfo{author}{\bibfnamefont{A.}~\bibnamefont{Knie}}, \bibnamefont{et~al.},
  \bibinfo{journal}{Chem. Rev.} \textbf{\bibinfo{volume}{120}},
  \bibinfo{pages}{11295} (\bibinfo{year}{2020}).

\bibitem[{\citenamefont{Marciniak et~al.}(2019)\citenamefont{Marciniak,
  Despré, Loriot, Karras, Herv\'e, Quintard, Catoire, Joblin, Constant, Kuleff
  et~al.}}]{Mar19}
\bibinfo{author}{\bibfnamefont{A.}~\bibnamefont{Marciniak}},
  \bibinfo{author}{\bibfnamefont{V.}~\bibnamefont{Despré}},
  \bibinfo{author}{\bibfnamefont{V.}~\bibnamefont{Loriot}},
  \bibinfo{author}{\bibfnamefont{G.}~\bibnamefont{Karras}},
  \bibinfo{author}{\bibfnamefont{M.}~\bibnamefont{Herv\'e}},
  \bibinfo{author}{\bibfnamefont{L.}~\bibnamefont{Quintard}},
  \bibinfo{author}{\bibfnamefont{F.}~\bibnamefont{Catoire}},
  \bibinfo{author}{\bibfnamefont{C.}~\bibnamefont{Joblin}},
  \bibinfo{author}{\bibfnamefont{E.}~\bibnamefont{Constant}},
  \bibinfo{author}{\bibfnamefont{A.}~\bibnamefont{Kuleff}},
  \bibnamefont{et~al.}, \bibinfo{journal}{Nat. Commun.}
  \textbf{\bibinfo{volume}{10}}, \bibinfo{pages}{337} (\bibinfo{year}{2019}).

\bibitem[{\citenamefont{Marroux et~al.}(2020)\citenamefont{Marroux, Fidler,
  Ghosh, Kobayashi, Gokhberg, Kuleff, Leone, and Neumark}}]{Mar20}
\bibinfo{author}{\bibfnamefont{H.~J.~B.} \bibnamefont{Marroux}},
  \bibinfo{author}{\bibfnamefont{A.~P.} \bibnamefont{Fidler}},
  \bibinfo{author}{\bibfnamefont{A.}~\bibnamefont{Ghosh}},
  \bibinfo{author}{\bibfnamefont{Y.}~\bibnamefont{Kobayashi}},
  \bibinfo{author}{\bibfnamefont{K.}~\bibnamefont{Gokhberg}},
  \bibinfo{author}{\bibfnamefont{A.~I.} \bibnamefont{Kuleff}},
  \bibinfo{author}{\bibfnamefont{S.~R.} \bibnamefont{Leone}}, \bibnamefont{and}
  \bibinfo{author}{\bibfnamefont{D.~M.} \bibnamefont{Neumark}},
  \bibinfo{journal}{Nat. Commun.} \textbf{\bibinfo{volume}{11}},
  \bibinfo{pages}{5810} (\bibinfo{year}{2020}).

\bibitem[{\citenamefont{Wahyutama et~al.}(2019)\citenamefont{Wahyutama, Sato,
  and Ishikawa}}]{Wah19}
\bibinfo{author}{\bibfnamefont{I.~S.} \bibnamefont{Wahyutama}},
  \bibinfo{author}{\bibfnamefont{T.}~\bibnamefont{Sato}}, \bibnamefont{and}
  \bibinfo{author}{\bibfnamefont{K.~L.} \bibnamefont{Ishikawa}},
  \bibinfo{journal}{Phys. Rev. A} \textbf{\bibinfo{volume}{99}},
  \bibinfo{pages}{063420} (\bibinfo{year}{2019}).

\bibitem[{\citenamefont{Austin et~al.}(2021)\citenamefont{Austin, Johnson,
  McGrath, Wood, Miseikis, Siegel, Hawkins, Harvey, Masin, Patchkovskii
  et~al.}}]{Aus21}
\bibinfo{author}{\bibfnamefont{D.~R.} \bibnamefont{Austin}},
  \bibinfo{author}{\bibfnamefont{A.~S.} \bibnamefont{Johnson}},
  \bibinfo{author}{\bibfnamefont{F.}~\bibnamefont{McGrath}},
  \bibinfo{author}{\bibfnamefont{D.}~\bibnamefont{Wood}},
  \bibinfo{author}{\bibfnamefont{L.}~\bibnamefont{Miseikis}},
  \bibinfo{author}{\bibfnamefont{T.}~\bibnamefont{Siegel}},
  \bibinfo{author}{\bibfnamefont{P.}~\bibnamefont{Hawkins}},
  \bibinfo{author}{\bibfnamefont{A.}~\bibnamefont{Harvey}},
  \bibinfo{author}{\bibfnamefont{Z.}~\bibnamefont{Masin}},
  \bibinfo{author}{\bibfnamefont{S.}~\bibnamefont{Patchkovskii}},
  \bibnamefont{et~al.}, \bibinfo{journal}{Sci. Rep.}
  \textbf{\bibinfo{volume}{11}}, \bibinfo{pages}{2485} (\bibinfo{year}{2021}).

\bibitem[{\citenamefont{Ruberti}(2019)}]{Rub19}
\bibinfo{author}{\bibfnamefont{M.}~\bibnamefont{Ruberti}},
  \bibinfo{journal}{Phys. Chem. Chem. Phys.} \textbf{\bibinfo{volume}{21}},
  \bibinfo{pages}{17584} (\bibinfo{year}{2019}).

\bibitem[{\citenamefont{Cederbaum et~al.}(1986)\citenamefont{Cederbaum, Domcke,
  Schirmer, and Niessen}}]{Ced86}
\bibinfo{author}{\bibfnamefont{L.~S.} \bibnamefont{Cederbaum}},
  \bibinfo{author}{\bibfnamefont{W.}~\bibnamefont{Domcke}},
  \bibinfo{author}{\bibfnamefont{J.}~\bibnamefont{Schirmer}}, \bibnamefont{and}
  \bibinfo{author}{\bibfnamefont{W.~V.} \bibnamefont{Niessen}},
  \bibinfo{journal}{Adv. Chem. Phys.} \textbf{\bibinfo{volume}{65}},
  \bibinfo{pages}{115} (\bibinfo{year}{1986}).

\bibitem[{\citenamefont{Cederbaum and Zobeley}(1999)}]{Ced99}
\bibinfo{author}{\bibfnamefont{L.}~\bibnamefont{Cederbaum}} \bibnamefont{and}
  \bibinfo{author}{\bibfnamefont{J.}~\bibnamefont{Zobeley}},
  \bibinfo{journal}{Chem. Phys. Lett.} \textbf{\bibinfo{volume}{307}},
  \bibinfo{pages}{205} (\bibinfo{year}{1999}).

\bibitem[{\citenamefont{Weinkauf et~al.}(1996)\citenamefont{Weinkauf, Schanen,
  Metsala, Schlag, Buergle, and Kessler}}]{Wei96}
\bibinfo{author}{\bibfnamefont{R.}~\bibnamefont{Weinkauf}},
  \bibinfo{author}{\bibfnamefont{P.}~\bibnamefont{Schanen}},
  \bibinfo{author}{\bibfnamefont{A.}~\bibnamefont{Metsala}},
  \bibinfo{author}{\bibfnamefont{E.~W.} \bibnamefont{Schlag}},
  \bibinfo{author}{\bibfnamefont{M.}~\bibnamefont{Buergle}}, \bibnamefont{and}
  \bibinfo{author}{\bibfnamefont{H.}~\bibnamefont{Kessler}},
  \bibinfo{journal}{J. Phys. Chem.} \textbf{\bibinfo{volume}{100}},
  \bibinfo{pages}{18567} (\bibinfo{year}{1996}).

\bibitem[{\citenamefont{Remacle et~al.}(1999)\citenamefont{Remacle, Levine,
  Schlag, and Weinkauf}}]{Rem99}
\bibinfo{author}{\bibfnamefont{F.}~\bibnamefont{Remacle}},
  \bibinfo{author}{\bibfnamefont{R.~D.} \bibnamefont{Levine}},
  \bibinfo{author}{\bibfnamefont{E.~W.} \bibnamefont{Schlag}},
  \bibnamefont{and} \bibinfo{author}{\bibfnamefont{R.}~\bibnamefont{Weinkauf}},
  \bibinfo{journal}{J. Phys. Chem. A} \textbf{\bibinfo{volume}{103}},
  \bibinfo{pages}{10149} (\bibinfo{year}{1999}).

\bibitem[{\citenamefont{Remacle and Levine}(2006)}]{Rem06}
\bibinfo{author}{\bibfnamefont{F.}~\bibnamefont{Remacle}} \bibnamefont{and}
  \bibinfo{author}{\bibfnamefont{R.~D.} \bibnamefont{Levine}},
  \bibinfo{journal}{Proc. Natl. Acad. Sci.} \textbf{\bibinfo{volume}{103}},
  \bibinfo{pages}{6793} (\bibinfo{year}{2006}).

\bibitem[{\citenamefont{Kuleff and Cederbaum}(2011)}]{Kul11}
\bibinfo{author}{\bibfnamefont{A.~I.} \bibnamefont{Kuleff}} \bibnamefont{and}
  \bibinfo{author}{\bibfnamefont{L.~S.} \bibnamefont{Cederbaum}},
  \bibinfo{journal}{Phys. Rev. Lett.} \textbf{\bibinfo{volume}{106}},
  \bibinfo{pages}{053001} (\bibinfo{year}{2011}).

\bibitem[{\citenamefont{Kuleff et~al.}(2016)\citenamefont{Kuleff, Kryzhevoi,
  Pernpointner, and Cederbaum}}]{Kul16}
\bibinfo{author}{\bibfnamefont{A.~I.} \bibnamefont{Kuleff}},
  \bibinfo{author}{\bibfnamefont{N.~V.} \bibnamefont{Kryzhevoi}},
  \bibinfo{author}{\bibfnamefont{M.}~\bibnamefont{Pernpointner}},
  \bibnamefont{and} \bibinfo{author}{\bibfnamefont{L.~S.}
  \bibnamefont{Cederbaum}}, \bibinfo{journal}{Phys. Rev. Lett.}
  \textbf{\bibinfo{volume}{117}}, \bibinfo{pages}{093002}
  (\bibinfo{year}{2016}).

\bibitem[{\citenamefont{Golubev et~al.}(2021)\citenamefont{Golubev,
  Van\'i\v{c}ek, and Kuleff}}]{Gol21}
\bibinfo{author}{\bibfnamefont{N.~V.} \bibnamefont{Golubev}},
  \bibinfo{author}{\bibfnamefont{J.}~\bibnamefont{Van\'i\v{c}ek}},
  \bibnamefont{and} \bibinfo{author}{\bibfnamefont{A.~I.}
  \bibnamefont{Kuleff}}, \bibinfo{journal}{Phys. Rev. Lett.}
  \textbf{\bibinfo{volume}{127}}, \bibinfo{pages}{123001}
  (\bibinfo{year}{2021}).

\bibitem[{\citenamefont{Gonçalves et~al.}(2021)\citenamefont{Gonçalves,
  Levine, and Remacle}}]{Gon21}
\bibinfo{author}{\bibfnamefont{C.~E.~M.} \bibnamefont{Gonçalves}},
  \bibinfo{author}{\bibfnamefont{R.~D.} \bibnamefont{Levine}},
  \bibnamefont{and} \bibinfo{author}{\bibfnamefont{F.}~\bibnamefont{Remacle}},
  \bibinfo{journal}{Phys. Chem. Chem. Phys.} \textbf{\bibinfo{volume}{23}},
  \bibinfo{pages}{12051} (\bibinfo{year}{2021}).

\bibitem[{\citenamefont{Khalili et~al.}(2021)\citenamefont{Khalili, Vafaee, and
  Shokri}}]{Kha21}
\bibinfo{author}{\bibfnamefont{F.}~\bibnamefont{Khalili}},
  \bibinfo{author}{\bibfnamefont{M.}~\bibnamefont{Vafaee}}, \bibnamefont{and}
  \bibinfo{author}{\bibfnamefont{B.}~\bibnamefont{Shokri}},
  \bibinfo{journal}{Phys. Chem. Chem. Phys.} \textbf{\bibinfo{volume}{23}},
  \bibinfo{pages}{23005} (\bibinfo{year}{2021}).

\bibitem[{\citenamefont{Runge and Gross}(1984)}]{Run84}
\bibinfo{author}{\bibfnamefont{E.}~\bibnamefont{Runge}} \bibnamefont{and}
  \bibinfo{author}{\bibfnamefont{E.~K.~U.} \bibnamefont{Gross}},
  \bibinfo{journal}{Phys. Rev. Lett.} \textbf{\bibinfo{volume}{52}},
  \bibinfo{pages}{997} (\bibinfo{year}{1984}).

\bibitem[{\citenamefont{Goedecker et~al.}(1996)\citenamefont{Goedecker, Teter,
  and Hutter}}]{Goe96}
\bibinfo{author}{\bibfnamefont{S.}~\bibnamefont{Goedecker}},
  \bibinfo{author}{\bibfnamefont{M.}~\bibnamefont{Teter}}, \bibnamefont{and}
  \bibinfo{author}{\bibfnamefont{J.}~\bibnamefont{Hutter}},
  \bibinfo{journal}{Phys. Rev. B} \textbf{\bibinfo{volume}{54}},
  \bibinfo{pages}{1703} (\bibinfo{year}{1996}).

\bibitem[{\citenamefont{Vidal et~al.}(2010)\citenamefont{Vidal, Wang, Dinh,
  Reinhard, and Suraud}}]{Vid10a}
\bibinfo{author}{\bibfnamefont{S.}~\bibnamefont{Vidal}},
  \bibinfo{author}{\bibfnamefont{Z.~P.} \bibnamefont{Wang}},
  \bibinfo{author}{\bibfnamefont{P.~M.} \bibnamefont{Dinh}},
  \bibinfo{author}{\bibfnamefont{P.-G.} \bibnamefont{Reinhard}},
  \bibnamefont{and} \bibinfo{author}{\bibfnamefont{E.}~\bibnamefont{Suraud}},
  \bibinfo{journal}{J. Phys. B} \textbf{\bibinfo{volume}{43}},
  \bibinfo{pages}{165102} (\bibinfo{year}{2010}).

\bibitem[{\citenamefont{Dinh et~al.}(2012)\citenamefont{Dinh, Vidal, Reinhard,
  and Suraud}}]{Din12b}
\bibinfo{author}{\bibfnamefont{P.~M.} \bibnamefont{Dinh}},
  \bibinfo{author}{\bibfnamefont{S.}~\bibnamefont{Vidal}},
  \bibinfo{author}{\bibfnamefont{P.-G.} \bibnamefont{Reinhard}},
  \bibnamefont{and} \bibinfo{author}{\bibfnamefont{E.}~\bibnamefont{Suraud}},
  \bibinfo{journal}{New J. Phys.} \textbf{\bibinfo{volume}{14}},
  \bibinfo{pages}{063015} (\bibinfo{year}{2012}).

\bibitem[{\citenamefont{Gao et~al.}(2015)\citenamefont{Gao, Wopperer, Dinh,
  Suraud, and Reinhard}}]{Wop15b}
\bibinfo{author}{\bibfnamefont{C.-Z.} \bibnamefont{Gao}},
  \bibinfo{author}{\bibfnamefont{P.}~\bibnamefont{Wopperer}},
  \bibinfo{author}{\bibfnamefont{P.~M.} \bibnamefont{Dinh}},
  \bibinfo{author}{\bibfnamefont{E.}~\bibnamefont{Suraud}}, \bibnamefont{and}
  \bibinfo{author}{\bibfnamefont{P.-G.} \bibnamefont{Reinhard}},
  \bibinfo{journal}{J. Phys. B} \textbf{\bibinfo{volume}{48}},
  \bibinfo{pages}{105102} (\bibinfo{year}{2015}).

\bibitem[{\citenamefont{Reinhard and Suraud}(2015)}]{Rei15d}
\bibinfo{author}{\bibfnamefont{P.-G.} \bibnamefont{Reinhard}} \bibnamefont{and}
  \bibinfo{author}{\bibfnamefont{E.}~\bibnamefont{Suraud}},
  \bibinfo{journal}{Ann. Phys. (N.Y.)} \textbf{\bibinfo{volume}{354}},
  \bibinfo{pages}{183} (\bibinfo{year}{2015}).

\bibitem[{\citenamefont{Dreizler and Gross}(1990)}]{Dre90}
\bibinfo{author}{\bibfnamefont{R.~M.} \bibnamefont{Dreizler}} \bibnamefont{and}
  \bibinfo{author}{\bibfnamefont{E.~K.~U.} \bibnamefont{Gross}},
  \emph{\bibinfo{title}{Density Functional Theory: An Approach to the \\Quantum
  Many-Body Problem}} (\bibinfo{publisher}{Springer-Verlag},
  \bibinfo{address}{Berlin}, \bibinfo{year}{1990}).

\bibitem[{\citenamefont{Perdew and Wang}(1992)}]{Per92}
\bibinfo{author}{\bibfnamefont{J.~P.} \bibnamefont{Perdew}} \bibnamefont{and}
  \bibinfo{author}{\bibfnamefont{Y.}~\bibnamefont{Wang}},
  \bibinfo{journal}{Phys. Rev. B} \textbf{\bibinfo{volume}{45}},
  \bibinfo{pages}{13244} (\bibinfo{year}{1992}).

\bibitem[{\citenamefont{Perdew and Zunger}(1981)}]{Per81}
\bibinfo{author}{\bibfnamefont{J.~P.} \bibnamefont{Perdew}} \bibnamefont{and}
  \bibinfo{author}{\bibfnamefont{A.}~\bibnamefont{Zunger}},
  \bibinfo{journal}{Phys. Rev. B} \textbf{\bibinfo{volume}{23}},
  \bibinfo{pages}{5048} (\bibinfo{year}{1981}).

\bibitem[{\citenamefont{Fermi and Amaldi}(1934)}]{Fer34}
\bibinfo{author}{\bibfnamefont{E.}~\bibnamefont{Fermi}} \bibnamefont{and}
  \bibinfo{author}{\bibfnamefont{E.}~\bibnamefont{Amaldi}},
  \bibinfo{journal}{Accad. Ital. Rome} \textbf{\bibinfo{volume}{6}},
  \bibinfo{pages}{117} (\bibinfo{year}{1934}).

\bibitem[{\citenamefont{Legrand et~al.}(2002)\citenamefont{Legrand, Suraud, and
  Reinhard}}]{Leg02a}
\bibinfo{author}{\bibfnamefont{C.}~\bibnamefont{Legrand}},
  \bibinfo{author}{\bibfnamefont{E.}~\bibnamefont{Suraud}}, \bibnamefont{and}
  \bibinfo{author}{\bibfnamefont{P.-G.} \bibnamefont{Reinhard}},
  \bibinfo{journal}{J. Phys. B} \textbf{\bibinfo{volume}{35}},
  \bibinfo{pages}{1115} (\bibinfo{year}{2002}).

\bibitem[{\citenamefont{Kl\"upfel et~al.}(2013)\citenamefont{Kl\"upfel, Dinh,
  Reinhard, and Suraud}}]{Klu13a}
\bibinfo{author}{\bibfnamefont{P.}~\bibnamefont{Kl\"upfel}},
  \bibinfo{author}{\bibfnamefont{P.~M.} \bibnamefont{Dinh}},
  \bibinfo{author}{\bibfnamefont{P.-G.} \bibnamefont{Reinhard}},
  \bibnamefont{and} \bibinfo{author}{\bibfnamefont{E.}~\bibnamefont{Suraud}},
  \bibinfo{journal}{Phys. Rev. A} \textbf{\bibinfo{volume}{88}},
  \bibinfo{pages}{052501} (\bibinfo{year}{2013}).

\bibitem[{\citenamefont{Reinhard and Suraud}(2021)}]{Rei21}
\bibinfo{author}{\bibfnamefont{P.-G.} \bibnamefont{Reinhard}} \bibnamefont{and}
  \bibinfo{author}{\bibfnamefont{E.}~\bibnamefont{Suraud}},
  \bibinfo{journal}{Theoret. Chem. Acc.} \textbf{\bibinfo{volume}{140}},
  \bibinfo{pages}{63} (\bibinfo{year}{2021}).

\bibitem[{\citenamefont{Ullrich}(2000)}]{Ull00b}
\bibinfo{author}{\bibfnamefont{C.~A.} \bibnamefont{Ullrich}},
  \bibinfo{journal}{J. Mol. Struct. (THEOCHEM)}
  \textbf{\bibinfo{volume}{501-502}}, \bibinfo{pages}{315}
  (\bibinfo{year}{2000}).

\bibitem[{\citenamefont{Reinhard et~al.}(2006)\citenamefont{Reinhard,
  Stevenson, Almehed, Maruhn, and Strayer}}]{Rei06c}
\bibinfo{author}{\bibfnamefont{P.-G.} \bibnamefont{Reinhard}},
  \bibinfo{author}{\bibfnamefont{P.~D.} \bibnamefont{Stevenson}},
  \bibinfo{author}{\bibfnamefont{D.}~\bibnamefont{Almehed}},
  \bibinfo{author}{\bibfnamefont{J.~A.} \bibnamefont{Maruhn}},
  \bibnamefont{and} \bibinfo{author}{\bibfnamefont{M.~R.}
  \bibnamefont{Strayer}}, \bibinfo{journal}{Phys. Rev. E}
  \textbf{\bibinfo{volume}{73}}, \bibinfo{pages}{036709}
  (\bibinfo{year}{2006}).

\bibitem[{\citenamefont{Dinh et~al.}(2018)\citenamefont{Dinh, Lacombe,
  Reinhard, Suraud, and Vincendon}}]{Din18}
\bibinfo{author}{\bibfnamefont{P.~M.} \bibnamefont{Dinh}},
  \bibinfo{author}{\bibfnamefont{L.}~\bibnamefont{Lacombe}},
  \bibinfo{author}{\bibfnamefont{P.-G.} \bibnamefont{Reinhard}},
  \bibinfo{author}{\bibfnamefont{E.}~\bibnamefont{Suraud}}, \bibnamefont{and}
  \bibinfo{author}{\bibfnamefont{M.}~\bibnamefont{Vincendon}},
  \bibinfo{journal}{Eur. Phys. J B} \textbf{\bibinfo{volume}{91}},
  \bibinfo{pages}{246} (\bibinfo{year}{2018}).

\bibitem[{\citenamefont{Dinh et~al.}(2022)\citenamefont{Dinh, Vincendon,
  Coppens, Suraud, and Reinhard}}]{Din22aR}
\bibinfo{author}{\bibfnamefont{P.~M.} \bibnamefont{Dinh}},
  \bibinfo{author}{\bibfnamefont{M.}~\bibnamefont{Vincendon}},
  \bibinfo{author}{\bibfnamefont{F.}~\bibnamefont{Coppens}},
  \bibinfo{author}{\bibfnamefont{E.}~\bibnamefont{Suraud}}, \bibnamefont{and}
  \bibinfo{author}{\bibfnamefont{P.-G.} \bibnamefont{Reinhard}},
  \bibinfo{journal}{Comp. Phys. Comm.} \textbf{\bibinfo{volume}{270}},
  \bibinfo{pages}{108155} (\bibinfo{year}{2022}).

\bibitem[{\citenamefont{Calvayrac et~al.}(2000)\citenamefont{Calvayrac,
  Reinhard, Suraud, and Ullrich}}]{Cal00}
\bibinfo{author}{\bibfnamefont{F.}~\bibnamefont{Calvayrac}},
  \bibinfo{author}{\bibfnamefont{P.-G.} \bibnamefont{Reinhard}},
  \bibinfo{author}{\bibfnamefont{E.}~\bibnamefont{Suraud}}, \bibnamefont{and}
  \bibinfo{author}{\bibfnamefont{C.~A.} \bibnamefont{Ullrich}},
  \bibinfo{journal}{Phys. Rep.} \textbf{\bibinfo{volume}{337}},
  \bibinfo{pages}{493} (\bibinfo{year}{2000}).

\bibitem[{\citenamefont{Calvayrac et~al.}(1997)\citenamefont{Calvayrac,
  Reinhard, and Suraud}}]{Cal97b}
\bibinfo{author}{\bibfnamefont{F.}~\bibnamefont{Calvayrac}},
  \bibinfo{author}{\bibfnamefont{P.-G.} \bibnamefont{Reinhard}},
  \bibnamefont{and} \bibinfo{author}{\bibfnamefont{E.}~\bibnamefont{Suraud}},
  \bibinfo{journal}{Ann. Phys. (N.Y.)} \textbf{\bibinfo{volume}{255}},
  \bibinfo{pages}{125} (\bibinfo{year}{1997}).

\bibitem[{\citenamefont{Pohl et~al.}(2000)\citenamefont{Pohl, Reinhard, and
  Suraud}}]{Poh00}
\bibinfo{author}{\bibfnamefont{A.}~\bibnamefont{Pohl}},
  \bibinfo{author}{\bibfnamefont{P.-G.} \bibnamefont{Reinhard}},
  \bibnamefont{and} \bibinfo{author}{\bibfnamefont{E.}~\bibnamefont{Suraud}},
  \bibinfo{journal}{Phys. Rev. Lett.} \textbf{\bibinfo{volume}{84}},
  \bibinfo{pages}{5090} (\bibinfo{year}{2000}).

\bibitem[{\citenamefont{Pohl et~al.}(2001)\citenamefont{Pohl, Reinhard, and
  Suraud}}]{Poh01}
\bibinfo{author}{\bibfnamefont{A.}~\bibnamefont{Pohl}},
  \bibinfo{author}{\bibfnamefont{P.-G.} \bibnamefont{Reinhard}},
  \bibnamefont{and} \bibinfo{author}{\bibfnamefont{E.}~\bibnamefont{Suraud}},
  \bibinfo{journal}{J. Phys. B} \textbf{\bibinfo{volume}{34}},
  \bibinfo{pages}{4969} (\bibinfo{year}{2001}).

\bibitem[{\citenamefont{Wopperer et~al.}(2015)\citenamefont{Wopperer, Dinh,
  Reinhard, and Suraud}}]{Wop15aR}
\bibinfo{author}{\bibfnamefont{P.}~\bibnamefont{Wopperer}},
  \bibinfo{author}{\bibfnamefont{P.~M.} \bibnamefont{Dinh}},
  \bibinfo{author}{\bibfnamefont{P.-G.} \bibnamefont{Reinhard}},
  \bibnamefont{and} \bibinfo{author}{\bibfnamefont{E.}~\bibnamefont{Suraud}},
  \bibinfo{journal}{Phys. Rep.} \textbf{\bibinfo{volume}{562}},
  \bibinfo{pages}{1} (\bibinfo{year}{2015}).

\bibitem[{\citenamefont{Dinh et~al.}(2013)\citenamefont{Dinh, Romaniello,
  Reinhard, and Suraud}}]{Din13c}
\bibinfo{author}{\bibfnamefont{P.~M.} \bibnamefont{Dinh}},
  \bibinfo{author}{\bibfnamefont{P.}~\bibnamefont{Romaniello}},
  \bibinfo{author}{\bibfnamefont{P.-G.} \bibnamefont{Reinhard}},
  \bibnamefont{and} \bibinfo{author}{\bibfnamefont{E.}~\bibnamefont{Suraud}},
  \bibinfo{journal}{Phys. Rev. A} \textbf{\bibinfo{volume}{87}},
  \bibinfo{pages}{032514} (\bibinfo{year}{2013}).

\bibitem[{\citenamefont{Dundas}(2012)}]{dundas:2012}
\bibinfo{author}{\bibfnamefont{D.}~\bibnamefont{Dundas}}, \bibinfo{journal}{J.
  Chem. Phys.} \textbf{\bibinfo{volume}{136}}, \bibinfo{pages}{194303}
  (\bibinfo{year}{2012}).

\bibitem[{\citenamefont{Haken}(1983)}]{Hak83aB}
\bibinfo{author}{\bibfnamefont{H.}~\bibnamefont{Haken}},
  \emph{\bibinfo{title}{{Laser Theory}}} (\bibinfo{publisher}{Springer},
  \bibinfo{address}{Berlin}, \bibinfo{year}{1983}).

\bibitem[{\citenamefont{Haken}(1991)}]{Hak91aB}
\bibinfo{author}{\bibfnamefont{H.}~\bibnamefont{Haken}},
  \emph{\bibinfo{title}{{Introduction to laser physics}}}
  (\bibinfo{publisher}{Springer}, \bibinfo{address}{Berlin},
  \bibinfo{year}{1991}).

\end{thebibliography}

\end{document}